\documentclass[preprint,aps]{revtex4}
\usepackage{amssymb,epsf}
\usepackage{amsmath}
\usepackage{amsfonts}
\begin{document}

\title{Counterterm Method in Lovelock Theory and Horizonless Solutions in Dimensionally Continued Gravity}
\author{M. H. Dehghani$^{1,2}$\footnote{email address:
mhd@shirazu.ac.ir}, N. Bostani$^{1}$ and A. Sheykhi$^{1}$}
\affiliation{$^1$Physics Department and Biruni Observatory,
College of Sciences, Shiraz University, Shiraz 71454, Iran\\
$^2$Research Institute for Astrophysics and Astronomy of Maragha
(RIAAM), Maragha, Iran}

\begin{abstract}
In this paper we, first, generalize the quasilocal definition of
the stress energy tensor of Einstein gravity to the case of
Lovelock gravity, by introducing the tensorial form of surface
terms that make the action well-defined. We also introduce the
boundary counterterm that removes the divergences of the action
and the conserved quantities of the solutions of Lovelock gravity
with flat boundary at constant $t$ and $r$. Second, we obtain the
metric of spacetimes generated by brane sources in dimensionally
continued gravity through the use of Hamiltonian formalism, and
show that these solutions have no curvature singularity and no
horizons, but have conic singularity. We show that these
asymptotically AdS spacetimes which contain two fundamental
constants are complete. Finally we compute the conserved
quantities of these solutions through the use of the counterterm
method introduced in the first part of the paper.
\end{abstract}

\maketitle

\section{Introduction}

A natural generalization of general relativity in higher
dimensional spacetimes with the assumption of Einstein -- that the
left hand side of the field equations is the most general
symmetric conserved tensor containing no more than second
derivatives of the metric -- is Lovelock theory. Lovelock
\cite{Lov} found the most general symmetric conserved tensor
satisfying this property. The resultant tensor is nonlinear in the
Riemann tensor and differs from the Einstein tensor only if the
spacetime has more than 4 dimensions. Since the Lovelock tensor
contains metric derivatives no higher than second order, the
quantization of the linearized Lovelock theory is ghost-free
\cite{Zw}.

Our first aim in this paper is to generalize the definition of the
quasilocal stress energy tensor for computing the conserved quantities of a
solution of Lovelock gravity. The concepts of action and energy-momentum
play central roles in gravity. However there is no good local notion of
energy for a gravitating system. A quasilocal definition of the energy and
conserved quantities for Einstein gravity can be found in \cite{BY}. They
define the quasilocal stress energy tensor through the use of the
well-defined gravitational action of Einstein gravity with the surface term
of Gibbons and Hawking \cite{Gib}. Therefore the first step is to find the
surface terms for the action of Lovelock gravity that make the action
well-defined. These surface terms were introduced by Myers in terms of
differential forms \cite{Myers}. Here, we write down the tensorial form of
the surface terms for Lovelock gravity, and then introduce the stress energy
tensor via the quasilocal formalism. The explicit form of these surface
terms for second and third order Lovelock gravity have been written in Refs.
\cite{Dav} and \cite{DM1} respectively. Of course, as in the case of
Einstein gravity, the action and conserved quantities diverge when the
boundary goes to infinity \cite{BY}. One way of eliminating these
divergences is through the use of background subtraction \cite{BY,BCM,BMann}%
, in which the boundary surface is embedded in another (background)
spacetime, and all quasilocal quantities are computed with respect to this
background, incorporated into the theory by adding to the action the
extrinsic curvature of the embedded surface. Such a procedure causes the
resulting physical quantities to depend on the choice of reference
background; furthermore, it is not possible in general to embed the boundary
surface into a background spacetime. For asymptotically AdS solutions, one
can instead deal with these divergences via the counterterm method inspired
by AdS/CFT correspondence \cite{Mal}. This conjecture, which relates the low
energy limit of string theory in asymptotically anti de-Sitter spacetime and
the quantum field theory on its boundary, has attracted a great deal of
attention in recent years. The equivalence between the two formulations
means that, at least in principle, one can obtain complete information on
one side of the duality by performing computation on the other side. A
dictionary translating between different quantities in the bulk gravity
theory and their counterparts on the boundary has emerged, including the
partition functions of both theories. In the present context this conjecture
furnishes a means for calculating the action and conserved quantities
intrinsically without reliance on any reference spacetime \cite{Sken,BK} by
adding additional terms on the boundary that are curvature invariants of the
induced metric. Although there may exist a very large number of possible
invariants one could add in a given dimension, only a finite number of them
are nonvanishing as the boundary is taken to infinity. Its many applications
include computations of conserved quantities for black holes with rotation,
NUT charge, various topologies, rotating black strings with zero curvature
horizons and rotating higher genus black branes \cite{EJM,Od1,Deh1}.
Although the counterterm method applies for the case of a specially infinite
boundary, it was also employed for the computation of the conserved and
thermodynamic quantities in the case of a finite boundary \cite{DM2}. All of
these works are limited to Einstein gravity. Here we apply the counterterm
method to the case of the solutions of dimensionally continued gravity. At
any given dimension there are only finitely many counterterms that one can
write down that do not vanish at infinity. This does not depend upon what
the bulk theory is -- i.e. whether or not it is Einstein, Gauss-Bonnet, or
Lovelock gravity. Indeed, for asymptotically AdS solutions, the boundary
counterterms that cancel divergences in Einstein Gravity should also cancel
divergences in Lovelock gravity. The coefficients will be different and
depends on $\Lambda $\ and Lovelock coefficients as we will see this for the
volume term in the flat boundary case below. An alternative method of
computing the finite action and conserved quantities has also been
considered in \cite{Ol} for Lovelock gravity.

Our second aim in this paper is to obtain asymptotically anti de Sitter
(AdS) horizonless solution of dimensionally continued gravity and
investigate their properties. Far from being just a mathematical curiosity,
Lovelock gravity as well as higher curvature theories in general, have
received in the recent past a renewed interest motivated by the hope of
learning something about the nature of quantum gravity. In particular, exact
static spherically symmetric black hole solutions of second order Lovelock
gravity have been found in Ref. \cite{Des}, and of the Gauss-Bonnet-Maxwell
and Born-Infeld-Gauss-Bonnet models in Ref. \cite{Wil1}. The thermodynamics
of the uncharged static spherically black hole solutions has been considered
in \cite{MS}, of solutions with nontrivial topology in \cite{Cai} and of
charged solutions in \cite{Wil1,Od2}. All of these known solutions in
Gauss-Bonnet gravity are static. Not long ago one of us has introduced two
new classes of rotating solutions of second order Lovelock gravity and
investigate their thermodynamics \cite{Deh2}, and made the first attempt for
finding exact static and rotating solutions in third order Lovelock gravity
with the quartic terms \cite{Deh3,DM1}. Very recently NUT charged black hole
solutions of Gauss-Bonnet gravity\ and Gauss-Bonnet-Maxwell gravity were
obtained \cite{DM3,Deh4}. Also the static spherically symmetric solutions of
the dimensionally continued gravity have been explored in Ref. \cite{Zan},
while black hole solutions with nontrivial topology in this theory have been
studied in Ref. \cite{Aros}. The thermodynamics of these solutions have been
investigated in Refs. \cite{Mun,Chr,Clu}.

In this paper we are dealing with the issue of the spacetimes generated by
brane sources in $D$-dimensional continued gravity that are horizonless and
have nontrivial external solutions. These kinds of solutions have been
investigated by many authors in four dimensions. Static uncharged
cylindrically symmetric solutions of Einstein gravity in four dimensions
were considered in \cite{Levi}. Similar static solutions in the context of
cosmic string theory were found in \cite{Vil}. All of these solutions \cite
{Levi,Vil} are horizonless and have a conical geometry; they are everywhere
flat except at the location of the line source. The extension to include the
electromagnetic field has also been done \cite{Muk,Lem1}. Here we present
the $D$-dimensional solution in dimensionally continued gravity, and use the
counterterm method to compute the conserved quantities of the system.

The outline of our paper is as follows. In Sec. \ref{Lov}, we give the
tensorial form of the surface terms that make the action well-defined,
generalize the Brown York energy-momentum tensor for Lovelock gravity, and
introduce the counterterm method for calculating the finite action and
conserved quantities of solutions of Lovelock gravity with flat boundary. In
Sec. \ref{Sol} we give a brief review of Hamiltonian formalism and introduce
the $D$-dimensional asymptotically AdS horizonless solutions of
dimensionally continued gravity in odd and even dimensions. We also
investigate the properties of these solutions and compute the finite
conserved quantities of them. We finish our paper with some concluding
remarks.

\section{Lovelock Gravity and the Counterterm Method\label{Lov}}

We consider a $D$-dimensional spacetime manifold ${\cal M}$ with metric $%
g_{\mu \nu }$. We denote the timelike and spacelike boundaries of ${\cal M}$
by $\partial {\cal M}$ and $\Sigma $ respectively. The metric and the
extrinsic curvature of the timelike boundary $\partial {\cal M}$ are denoted
by $\gamma _{ab}$ and $\Theta _{ab}$, while those of the spacelike\
hypersurface ${\Sigma }$ are denoted by $h_{ij}$ and $K_{ij}$. In this $D$%
-dimensional spacetime, the most general action which keeps the field
equations of motion for the metric of second order, as the pure
Einstein-Hilbert action, is Lovelock action. This action is constructed from
the dimensionally extended Euler densities and can be written as
\begin{equation}
I_{G}=\kappa \int d^{D}x\sqrt{-g}\sum_{p=0}^{n}\alpha _{p}{\cal L}_{p}
\label{Lov1}
\end{equation}
where $n\equiv \lbrack (D-1)/2]$ and $[z]$ denotes the integer part of $z$, $%
\alpha _{p}$ is an arbitrary constant and ${\cal L}_{p}$ is the Euler
density of a $2p$-dimensional manifold
\begin{equation}
\mathcal{L}_{p}=\frac{1}{2^{p}}\delta _{\rho _{1}\sigma _{1}\cdots
\rho _{p}\sigma _{p}}^{\mu _{1}\nu _{1}\cdots \mu _{p}\nu
_{p}}R_{\mu _{1}\nu
_{1}}^{\phantom{\mu_1\nu_1}{\rho_1\sigma_1}}\cdots R_{\mu _{p}\nu _{p}}^{%
\phantom{\mu_k \nu_k}{\rho_p \sigma_p}}  \label{Lov2}
\end{equation}
In Eq. (\ref{Lov2}) $\delta _{\rho _{1}\sigma _{1}\cdots \rho _{p}\sigma
p}^{\mu _{1}\nu _{1}\cdots \mu p\nu _{p}}$ is the generalized totally
anti-symmetric Kronecker delta and $R_{\mu \nu }^{\phantom{\mu\nu}{\rho%
\sigma}}$ is the Riemann tensor of the Manifold ${\cal M}$. We note that in $%
D$ dimensions, all terms for which $p>[D/2]$ are total derivatives, and the
term $p=D/2$ is the Euler density. Consequently only terms for which $p<D/2$
contribute to the field equations.

The Einstein-Hilbert action (with $\alpha _{p}=0$ for $p\geq 2$) does not
have a well-defined variational principle, since one encounters a total
derivative that produces a surface integral involving the derivative of $%
\delta g_{\mu \nu }$ normal to the boundary $\partial {\cal M}$. These
normal derivative terms do not vanish by themselves, but are canceled by the
variation of the Gibbons-Hawking surface term \cite{Gib}
\begin{equation}
I_{b}^{(1)}=2\kappa \int_{\partial {\cal M}}d^{D-1}x\sqrt{-\gamma }\Theta
\label{Ib1}
\end{equation}
where $\gamma _{ab}$ is induced metric on the boundary $r={\rm const.}$ and $%
\Theta $ is the trace of extrinsic curvature of this boundary. The
main difference between higher derivative gravity and Einstein
gravity is that the surface term that renders the variational
principle well-behaved is much more complicated. However, the
surface terms that make the variational principle of Lovelock
gravity well-defined are known in terms of differential forms
\cite{Myers}. The tensorial form of these surface terms may be
written as

\begin{equation}
I_{b}=-2\kappa \int_{\partial {\cal M}}d^{D-1}x\sqrt{-\gamma }%
\sum_{p=0}^{n}\sum_{s=0}^{p-1}\frac{(-1)^{p-s}p\alpha _{p}}{2^{s}(2p-2s-1)}%
{\cal H}^{(p)}  \label{Ib}
\end{equation}
{\bf \ }where $\alpha _{p}$ is the Lovelock coefficients and ${\cal H}^{(p)}$
is
\begin{equation}
{\cal H}^{(p)}=\delta _{\,[b_{1}\ldots b_{2p-1}]}^{[a_{1}\ldots a_{2p-1}]}R_{%
\phantom{b_1b_1}{a_1a_2}}^{b_{1}b_{2}}\cdots R_{%
\phantom{b_{2s}}{a_{2s-1}a_{2s}}}^{b_{2s-1}b_{2s}}\Theta _{a1}^{b1}\cdots
\Theta _{a_{2p-1}}^{b_{2p-1}}  \label{Hp}
\end{equation}
In Eq. (\ref{Hp}) $R_{\phantom{cd}{cd}}^{ab}(g)$'s are the boundary
components of the Riemann tensor of the Manifold ${\cal M}$, which depend on
the velocities through the Gauss--Codazzi equations

\begin{equation}
R_{abcd}=\widehat{R}_{abcd}+\Theta _{ac}\Theta _{bd}-\Theta _{ad}\Theta _{bc}
\label{Codac}
\end{equation}
where $\widehat{R}_{abcd}(\gamma )$ are the components of the intrinsic
curvature tensor of the boundary. The explicit form of the second and third
surface terms of Eq. (\ref{Ib}) may be written as \cite{Dav,DM1}
\begin{eqnarray}
I_{b}^{(2)} &=&2\kappa \int_{\partial {\cal
M}}d^{D-1}x\sqrt{-\gamma }\left\{ 2\alpha _{2}\left(
J-2\widehat{G}_{ab}^{(1)}\Theta ^{ab}\right) \right.
\nonumber \\
&&\left. +3\alpha _{3}\left( P-2\widehat{G}_{ab}^{(2)}\Theta ^{ab}-12%
\widehat{R}_{ab}J^{ab}+2\widehat{R}J-4\Theta \widehat{R}_{abcd}\Theta
^{ac}\Theta ^{bd}-8\widehat{R}_{abcd}\Theta ^{ac}\Theta _{e}^{b}\Theta
^{ed}\right) \right\}
\end{eqnarray}
where $\widehat{G}_{ab}^{(1)}$ is the $n$-dimensional Einstein tensor of the
metric $\gamma _{ab}$, $J$ is the trace of
\begin{equation}
J_{ab}=\frac{1}{3}(2\Theta \Theta _{ac}\Theta _{b}^{c}+\Theta _{cd}\Theta
^{cd}\Theta _{ab}-2\Theta _{ac}\Theta ^{cd}\Theta _{db}-\Theta ^{2}\Theta
_{ab}),  \label{Jab}
\end{equation}
$\widehat{G}_{ab}^{(2)}$ is the second order Lovelock tensor for the
boundary metric $\gamma _{ab}$:
\begin{equation}
G_{a b}^{(2)}=2(\widehat{R}_{acde}\widehat{R}_{b}^{\phantom{b}cde}-2%
\widehat{R}_{acbd}\widehat{R}^{cd}-2\widehat{R}_{ac}R_{\phantom{c}%
b}^{c}+RR_{ab})-\frac{1}{2}(\widehat{R}_{cdef}\widehat{R}^{cdef}-4\widehat{R}%
_{cd}\widehat{R}^{cd}+\widehat{R}^{2})\gamma _{ab}  \label{Love2}
\end{equation}
and $P$ is the trace of
\begin{eqnarray}
P_{ab} &=&\frac{1}{5}\left\{ \left[ \Theta ^{4}-6\Theta ^{2}\Theta
^{cd}\Theta _{cd}+8\Theta \Theta _{cd}\Theta _{e}^{d}\Theta ^{ec}-6\Theta
_{cd}\Theta ^{de}\Theta _{ef}\Theta ^{fc}+3(\Theta _{cd}\Theta ^{cd})^{2}%
\right] \Theta _{ab}\right.  \nonumber \\
&&\left. -(4\Theta ^{3}-12\Theta \Theta _{ed}\Theta ^{ed}+8\Theta
_{de}\Theta _{f}^{e}\Theta ^{fd})\Theta _{ac}\Theta _{b}^{c}-24\Theta \Theta
_{ac}\Theta ^{cd}\Theta _{de}\Theta _{b}^{e}\right.  \nonumber \\
&&\left. +(12\Theta ^{2}-12\Theta _{ef}\Theta ^{ef})\Theta _{ac}\Theta
^{cd}\Theta _{db}+24\Theta _{ac}\Theta ^{cd}\Theta _{de}\Theta ^{ef}\Theta
_{bf}\right\}  \label{Pab}
\end{eqnarray}

In general $I=I_{G}+I_{b}$ is divergent when evaluated on solutions, as is
the Hamiltonian and other associated conserved quantities. In Einstein
gravity, one can remove the non logarithmic divergent terms in the action by
adding a counterterm action $I_{ct}$ which is a functional of the boundary
curvature invariants \cite{Kraus}. The issue of determination of boundary
counterterms with their coefficients for higher-order Lovelock theories is
at this point an open question. However for the case of a boundary with zero
curvature [$\widehat{R}_{abcd}(\gamma )=0$], it is quite straightforward.
This is because all curvature invariants are zero except for a constant, and
so the only possible boundary counterterm is one proportional to the volume
of the boundary regardless of the number of dimensions:

\begin{equation}
I_{ct}=2\kappa \lambda \alpha _{0}\int_{\partial {\cal M}_{\infty }}d^{D-1}x%
\sqrt{-\gamma }  \label{Ict}
\end{equation}
where $\lambda $ is a constant which should be chosen such that the
divergences of the action is removed.

Having the total finite action, one can use the quasilocal definition of
Brown and York \cite{BY} to construct a divergence free stress-energy tensor
as

\begin{equation}
T_{b}^{a}=-2\kappa \left\{ \lambda \alpha _{0}\gamma
_{b}^{a}+\sum_{p=0}^{n}\sum_{s=0}^{p-1}\frac{(-1)^{p-s}p\alpha _{p}}{%
2^{s}(2p-2s-1)}{\cal H}_{b}^{(p,s)a}\right\}  \label{Stres}
\end{equation}
where ${\cal H}_{\phantom{i}b}^{(p,s)a}$ is
\begin{equation}
\mathcal{H}_{\phantom{i}b}^{(p,s)a}=\delta _{\,[b_{1}\ldots
b_{2p-1}b]}^{[a_{1}\ldots a_{2p-1}a]}\widehat{R}_{\phantom{b_1b_2}{a_1a_2}%
}^{b_{1}b_{2}}\cdots \widehat{R}_{\phantom{b_{2s-1}b_{2s}}{a_{2s-1}a_{2s}}%
}^{b_{2s-1}b_{2s}}\Theta _{a_{2s+1}}^{b_{2s+1}}\cdots \Theta
_{a_{2p-1}}^{b_{2p-1}},  \label{Hps}
\end{equation}

\ To compute the conserved mass of the spacetime, one should choose a
spacelike surface ${\cal B}$ in $\partial {\cal M}$ with metric $\sigma _{ij}
$, and write the boundary metric in Arnowitt-Deser-Misner (ADM) form:
\[
\gamma _{ab}dx^{a}dx^{a}=-N^{2}dt^{2}+\sigma _{ij}\left( d\varphi
^{i}+N^{i}dt\right) \left( d\varphi ^{j}+N^{j}dt\right)
\]
where the coordinates $\varphi ^{i}$ are the angular variables
parameterizing the hypersurface of constant $r$ around the origin, and $N$
and $N^{i}$ are the lapse and shift functions respectively. When there is a
Killing vector field $\xi $ on the boundary, then the quasilocal conserved
quantities associated with the stress tensors of Eq. (\ref{Stres}) can be
written as
\begin{equation}
{\cal Q}(\xi )=\int_{{\cal B}}d^{D-2}\varphi \sqrt{\sigma }T_{ab}n^{a}\xi
^{b}  \label{charge}
\end{equation}
where $\sigma $ is the determinant of the metric $\sigma _{ab}$, $\xi $ and $%
n^{a}$ are the Killing vector field and the unit normal vector on the
boundary ${\cal B}$. In the context of counterterm method, the limit in
which the boundary ${\cal B}$ becomes infinite (${\cal B}_{\infty }$) is
taken, and the counterterm prescription ensures that the action and
conserved charges are finite. No embedding of the surface ${\cal B}$ in to a
reference of spacetime is required and the quantities which are computed are
intrinsic to the spacetimes.

\section{Horizonless Solutions in Dimensionally Continued Gravity\label{Sol}}

The dimensionally continued gravity is a special class of the Lovelock
gravity, in which the Lovelock coefficients are reduced to two by embedding
the Lorentz group $SO(D-1,1)$ into a larger AdS group $SO(D-1,2)$ \cite{Zan}%
. The remaining two fundamental constants are the gravitational and
cosmological constants. In odd dimensions it is possible to construct a
Lagrangian invariant under the anti-de Sitter group by making a certain
choice of the Lovelock coefficients, while it is not possible to construct a
non-trivial action principle invariant under $SO(D-1,2)$ and it is necessary
to break the symmetry down to the Lorentz group. Accordingly, Lovelock
gravity is separated into two distinct type of branches for odd and even
dimensions. In what follows, we will consider a particular choice of the
Lovelock coefficients given by
\begin{equation}
\alpha _{p}=\left\{
\begin{array}{ll}
(D-2p-1)!\left(
\begin{array}{c}
(D-1)/2 \\
p
\end{array}
\right) l^{2p-D} & \ \ \text{{\rm for odd}}\ D \\
(D-2p)!\left(
\begin{array}{c}
D/2 \\
p
\end{array}
\right) l^{2p-D} & \ \ \text{{\rm for} {\rm even}}\ D
\end{array}
\right.   \label{alpha}
\end{equation}
where $l$ is a length. For later convenience, the units are chosen such that
\begin{equation}
\kappa =\left\{
\begin{array}{ll}
-\frac{l^{D-2}}{2(D-3)!} & \ \ \text{{\rm for odd}}\ D \\
-\frac{l^{D-2}}{2D(D-3)!} & \ \ \text{{\rm for} {\rm even}}\ D
\end{array}
\right.   \label{kappa}
\end{equation}

In order to obtain simplified equations of motion, it is more convenient to
work in the Hamiltonian formalism. The Hamiltonian form of the action (\ref
{Lov1}) is discussed in \cite{TZ}. In that approach, the second order
formalism is used. The torsion tensor is set to zero and the connection is
solved in terms of the local frame and their derivatives. Just as in $D=4$,
the canonical coordinates are the spatial components of the metric $g_{ij}$,
and their conjugate momenta $\pi ^{ij}$. The time components $g_{0\mu }$ are
Lagrange multipliers associated with the generators of surface deformations,
${\cal H}_{\mu }=({\cal H},{\cal H}_{i})$. The action (\ref{Lov1}) takes the
form \cite{TZ}

\begin{equation}
I=\kappa \int (\pi ^{ij}\dot{h_{ij}}-N\,{\cal H}-N^{i}{\cal H}_{i})d^{%
\mbox{\tiny $\cal
D$}-1}xdt+B  \label{B1}
\end{equation}
where $N$, and $N^{i}$ are the lapse function and shift vectors in the
standard ADM decomposition of spacetime. The momenta $\pi _{j}^{i}$ and the
normal generator ${\cal H}$ are

\begin{equation}
\pi _{j}^{i}=-\frac{1}{4}\sqrt{-g}\sum_{p=0}^{n}\frac{\alpha _{p}}{2^{p}}%
\sum_{s=0}^{p-1}\frac{(-4)^{p-s}}{s![2(p-s)-1]!!}{\cal H}_{\phantom{i}%
j}^{(p)i}  \label{B2}
\end{equation}
where ${\cal H}_{\phantom{i}j}^{(p,s)i}$ is given by Eqs. (\ref{Hps}) which
is now evaluated on the spacelike hypersurface $\Sigma $ with the metric $%
h_{ij}$. The normal generator ${\cal H}$ is given in term of the spatial
components of Riemann tensor of the spacetime as
\begin{equation}
{\cal H}=-\sqrt{h}\sum_{p=0}^{n}\frac{\alpha _{p}}{2^{p}}\delta
_{\,[j_{1}\ldots j_{2p}]}^{[i_{1}\ldots i_{2p}]}R_{\phantom{j_1j_1}{i_1i_2}%
}^{j_{1}j_{2}}\cdots
R_{\phantom{j_{2p}}{i_{2p-1}i_{2p}}}^{j_{2p-1}j_{2p}}
\end{equation}
In the above equation $R_{\phantom{ij}{kl}}^{ij}$ depend on the velocities
through the Gauss--Codazzi equations
\begin{equation}
R_{ijkl}=\tilde{R}_{ijkl}+K_{ik}K_{jl}-K_{il}K_{jk}
\end{equation}
where $\tilde{R}_{ijkl}(h)$ are the components of the intrinsic curvature
tensor of the boundary $\Sigma $. The generators of reparameterizations of
the surfaces $t={\rm const}$, ${\cal H}_{i}=-2D_{j}\pi _{\;i}^{j}$ do not
depend on the action but only on the transformation laws of $h_{ij}$ and $%
\pi ^{ij}$.

Here we consider the spacetimes generated by brane sources in $D$%
-dimensional spacetime that are horizonless and have nontrivial
external solutions. We will work with the following ansatz for the
metric:
\begin{equation}
ds^{2}=-N^{2}(\rho )dt^{2}+\frac{d\rho ^{2}}{F(\rho )}+l^{2}F(\rho )d\phi
^{2}+\frac{\rho ^{2}}{l^{2}}dX_{D-3}^{2}  \label{met1}
\end{equation}
where $dX_{D-3}^{2}=\sum_{i=0}^{D-3}{(dx^{i})}^{2}$ is the Euclidean metric
of $(D-3)$-dimensional submanifold. The parameter $l^{2}$ is appropriate
constant proportional to cosmological constant $\Lambda $. The functions $%
N(\rho )$ and $F(\rho )$ need to be determined. The motivation for this
metric gauge $[(g_{\rho \rho })^{-1}\varpropto g_{\varphi \varphi }]$
instead of the usual Schwarzschild gauge $[(g_{\rho \rho })^{-1}\varpropto
g_{tt}]$ comes from the fact that we are looking for a string solution with
conic singularity. For the metric (\ref{met1}), all the components of the
shift vector and extrinsic curvature are zero. The normal generator ${\cal H}
$ can be computed easily as
\begin{equation}
{\cal H}=-(D-3)!\frac{1}{l^{D-4}}\left[ \rho ^{D-1}\sum_{p=0}^{n}\frac{%
\alpha _{p}}{(D-2p-1)!}\left( \frac{-F(\rho )}{\rho ^{2}}\right) ^{p}\right]
^{\prime \prime }  \label{ham2}
\end{equation}
where double prime denotes second derivative with respect to $\rho $. Using
the coefficients (\ref{alpha}) and units (\ref{kappa}) one obtains the
action
\begin{equation}
I=-l^{D-3}\omega _{D-2}(t_{2}-t_{1})\int d\rho \ N(\rho )G^{\prime \prime
}[\rho ,F(\rho )]+B,  \label{action2}
\end{equation}
where $B$ stands for a surface term, and $G$ is

\begin{equation}
G[\rho ,F(\rho )]=\left\{
\begin{array}{ll}
\frac{1}{2}{}[(\rho /l)^{2}-F(\rho )]^{n} & \ \text{{\rm for odd}}\ D \\
\frac{\rho }{2l}[(\rho /l)^{2}-F(\rho )]^{n} & \ \ \text{{\rm for even}}\ D
\end{array}
\right.   \label{Gfunction}
\end{equation}
where $\omega _{D-2}$ is the volume of the $(D-2)$-dimensional hypersurface
of constant $t$ and $r$. Varying the action (\ref{action2}) with respect to $%
N(\rho )$ and $F(\rho )$, one obtains the equations of motion as
\begin{equation}
G^{\prime \prime }=0,\ \ \ \ N^{\prime \prime }=0,  \label{eqmo}
\end{equation}
with the solutions
\begin{eqnarray}
&&G[\rho ,F(\rho )]=c\rho +m \label{Fr}\\
&&N(\rho )=C_{1}\rho +C_{2}. \label{Nr}
\end{eqnarray}
where $m$, $c$, $C_{1}$ and $C_{2}$ are arbitrary integration
constant. In order to reconstruct the four-dimensional solution of
Ref. \cite{Lem1}, we assume $C_{2}=0$. Since $N(\rho ) $ is
dimensionless and any constant may be absorbed in $t$, therefore,
we can choose $C_{1}=l^{-1}$ without loss of generality. Using
Eqs. (\ref{Gfunction}) and (\ref{Fr}), the metric function $F(\rho
)$ may be obtained as
\begin{equation}
F(\rho )=\left\{
\begin{array}{ll}
\frac{\rho ^{2}}{l^{2}}-(2c\rho +2m)^{\frac{1}{n}} & \ \ \text{{\rm for odd}}%
\ D \\
\frac{\rho ^{2}}{l^{2}}-\left( 2lc+\frac{2lm}{\rho }\right) ^{\frac{1}{n}} &
\ \ \text{{\rm for even}}\ D
\end{array}
\right.
\end{equation}
In order to study the general structure of this solution, we first look for
curvature singularities. It is easy to show that the Kretschmann scalar $%
R_{\mu \nu \lambda \kappa }R^{\mu \nu \lambda \kappa }$ diverge at $\rho =0$
and therefore one might think that there is a curvature singularity located
at $\rho =0$. However, as we will see below, the spacetime will never
achieve $\rho =0$. The function $F(\rho )$ is negative for $\rho <r_{+}$ and
positive for $\rho >r_{+}$, where $r_{+}$ is the largest root of $F(\rho )=0$%
. Indeed, $g_{\rho \rho }$ and $g_{\phi \phi }$ are related by $F(\rho
)=g_{\rho \rho }^{-1}=l^{-2}g_{\phi \phi }$, and therefore when $g_{\rho
\rho }$ becomes negative (which occurs for$\rho <r_{+}$) so does $g_{\phi
\phi }$. This leads to apparent change of signature of the metric from $%
(D-2)+$ to $(D-3)+$ as one extend the spacetime to $\rho <r_{+}$. This
indicates that we are using an incorrect extension. To get rid of this
incorrect extension, we introduce the new radial coordinate $r$ as\newline
\begin{equation}
r^{2}=\rho ^{2}-r_{+}^{2}\Longrightarrow d\rho ^{2}=\frac{r^{2}}{%
r^{2}+r_{+}^{2}}dr^{2}
\end{equation}
With this new coordinate, the metric (\ref{met1}) is\newline
\begin{equation}
ds^{2}=-\frac{r^{2}+r_{+}^{2}}{l^{2}}dt^{2}+l^{2}F(r)d\phi ^{2}+\frac{r^{2}}{%
(r^{2}+r_{+}^{2})F(r)}dr^{2}+\frac{r^{2}+r_{+}^{2}}{l^{2}}dX_{D-3}^{2}
\label{met2}
\end{equation}
where the coordinate $r$ and $\phi $ assume the values $0\leqslant r<\infty $
and $0\leqslant \phi <2\pi \ $, and $F(r)$ is now given as
\begin{equation}
F(r)=\left\{
\begin{array}{ll}
\frac{r^{2}+r_{+}^{2}}{l^{2}}-(2c\sqrt{r^{2}+r_{+}^{2}}+2m)^{\frac{1}{n}} &
\ \ \text{{\rm for odd}}\ D \\
\frac{r^{2}+r_{+}^{2}}{l^{2}}-\left( 2lc+\frac{2lm}{\sqrt{r^{2}+r_{+}^{2}}}%
\right) ^{\frac{1}{n}} & \ \ \text{{\rm for even}}\ D
\end{array}
\right.   \label{F2}
\end{equation}

The function $F(r)$ given in Eq. (\ref{F2}) is positive in the whole
spacetime and is zero at $r=0$. Also note that the Kretschmann scalar does
not diverge in the range $0\leqslant r<\infty $. Therefore this spacetime
has no curvature singularities and no horizons. However, it has a conic
geometry and has a conical singularity at $r=0$. In fact, using a Taylor
expansion in the vicinity of $r=0$ the metric (\ref{met2}) becomes
\begin{equation}
ds^{2}=-\frac{r_{+}^{2}}{l^{2}}dt^{2}+\frac{r^{2}dr^{2}}{r_{+}^{2}\left[
\frac{r_{+}^{2}}{l^{2}}-\left( 2cr_{+}+2m\right) ^{\frac{1}{n}}\right] }%
+l^{2}\left[ \frac{r_{+}^{2}}{l^{2}}-\left( 2cr_{+}+2m\right) ^{\frac{1}{n}}%
\right] d\phi ^{2}+\frac{r_{+}^{2}}{l^{2}}dX_{D-3}^{2}  \label{metodd}
\end{equation}
and

\begin{equation}
ds^{2}=-\frac{r_{+}^{2}}{l^{2}}dt^{2}+\frac{r^{2}dr^{2}}{r_{+}^{2}\left[
\frac{r_{+}^{2}}{l^{2}}-\left( 2lc+\frac{2lm}{r_{+}}\right) ^{\frac{1}{n}}%
\right] }+l^{2}\left[ \frac{r_{+}^{2}}{l^{2}}-\left( 2lc+\frac{2lm}{r_{+}}%
\right) ^{\frac{1}{n}}\right] d\phi ^{2}+\frac{r_{+}^{2}}{l^{2}}dX_{D-3}^{2}
\label{meteven}
\end{equation}
for odd-dimensional and even-dimensional spacetimes respectively. Equations (%
\ref{metodd}) and (\ref{meteven}) clearly show that the spacetime has a
conical singularity at $r=0$ in any dimensions.

Of course, one may ask for the completeness of the spacetime with $r\geq 0$
\cite{Lem1,Hor}. It is easy to see that the spacetime described by Eq. (\ref
{met2}) is both null and timelike geodesically complete for $r\geq 0$. To do
this, one may show that every null or timelike geodesic starting from an
arbitrary point either can be extended to infinite values of the affine
parameter along the geodesic or will end on a singularity at $r=0$. Using
the geodesic equation, one obtains
\begin{eqnarray}
\dot{t} &=&\frac{l^{2}}{r^{2}+r_{+}^{2}}E,\hspace{0.5cm}\dot{x^{i}}=\frac{%
l^{2}}{r^{2}+r_{+}^{2}}P^{i},\hspace{0.5cm}\dot{\phi}=\frac{1}{l^{2}F(r)}L,
\\
r^{2}\dot{r}^{2} &=&(r^{2}+r_{+}^{2})F(r)\left[ \frac{l^{2}(E^{2}-{\bf P}%
^{2})}{r^{2}+r_{+}^{2}}-\eta \right] -\frac{r^{2}+r_{+}^{2}}{l^{2}}L^{2}
\label{Geo2}
\end{eqnarray}
where the overdot denotes the derivative with respect to an affine
parameter, and $\eta $ is zero for null geodesics and $+1$ for timelike
geodesics. $E$, $L$, and $P^{i}$'s are the conserved quantities associated
with the coordinates $t$, $\phi $, and $x^{i}$, respectively, and ${\bf P}%
^{2}=\sum_{i=1}^{n-2}(P^{i})^{2}$. Notice that $F(r)$ is always positive for
$r>0$ and zero for $r=0$.

First we consider the null geodesics ($\eta =0$). (i) If $E^{2}>{\bf P}^{2}$
the spiraling particles ($L>0$) coming from infinity have a turning point at
$r_{tp}>0$, while the nonspiraling particles ($L=0$) have a turning point at
$r_{tp}=0$. (ii) If $E^{2}={\bf P}^{2}$ and $L=0$, whatever the value of $r$%
, $\dot{r}$ and $\dot{\phi}$ vanish and therefore the null particles moves
on the $z$-axis. (iii) For $E^{2}={\bf P}^{2}$ and $L\neq 0$, and also for $%
E^{2}<{\bf P}^{2}$ and any values of $L$, there is no possible null geodesic.

Second, we analyze the timelike geodesics ($\eta =+1$). Timelike geodesics
is possible only if $l^{2}(E^{2}-{\bf P}^{2})>r_{+}^{2}$. In this case
spiraling ($L\neq 0$) timelike particles are bound between $r_{tp}^{a}$ and $%
r_{tp}^{b}$ given by
\begin{equation}
0<r_{tp}^{a}\leq r_{tp}^{b}<\sqrt{l^{2}(E^{2}-{\bf P}^{2})-r_{+}^{2}},
\end{equation}
while the turning points for the nonspiraling particles ($L=0$) are $%
r_{tp}^{1}=0$ and $r_{tp}^{2}=\sqrt{l^{2}(E^{2}-{\bf P}^{2})-r_{+}^{2}}$.

\subsection{Conserved Quantities}

Now we apply the counterterm method to compute the conserved quantities of
the solution (\ref{met2}). For the horizonless spacetime (\ref{met2}), the
Killing vector is $\xi =\partial /\partial t$ and therefore its associated
conserved charge is the total mass of the system enclosed by the boundary
given as
\begin{equation}
M=\int_{{\cal B}}d^{D-2}\varphi \sqrt{\sigma }T_{ab}n^{a}\xi ^{b}
\label{Mass}
\end{equation}
where $T_{ab}$ is the stress energy tensor (\ref{Stres}). It is a matter of
calculation to show that the mass per unit volume $\omega _{D-2}$ is
\[
M=m
\]

\section{CLOSING REMARKS}

In this paper, first, we wrote down the surface terms for Lovelock gravity
which make the action well-defined. This is achieved by generalizing the
Gibbons-Hawking surface term for Einstein gravity. We also generalized the
stress energy momentum tensor of Brown and York in Einstein gravity \cite{BY}
to the case of Lovelock gravity. As in the case of Einstein gravity, $I_{G}$%
, and $I_{b}$ of Eqs. (\ref{Lov1}) and (\ref{Ib}) are divergent when
evaluated on the solutions, as is the Hamiltonian and other associated
conserved quantities. In Einstein gravity, one can remove these divergent
terms in the action by adding a counterterm action $I_{ct}$ which is a
functional of the boundary curvature invariants \cite{Kraus}. Although the
counterterms are not known for Lovelock gravity, for the case of a boundary
with zero curvature [$\widehat{R}_{abcd}(\gamma )=0$], it is quite
straightforward. This is because all curvature invariants are zero except
for a constant, and so the only possible boundary counterterm is one
proportional to the volume of the boundary regardless of the number of
dimensions. We, therefore, introduced the counterterm which removed the
divergences of the action and conserved quantities of the solutions of
Lovelock gravity with zero curvature boundary.

Second, we considered the asymptotically AdS horizonless solutions in
dimensionally continued gravity. We found a new class of solutions in
dimensionally continued gravity through the use of Hamiltonian formalism
which has no curvature singularity and no horizons, but have conic
singularity at $r=0$. These horizonless solutions have two fundamental
constants which are the Newton's and cosmological constants. We showed that
these spacetimes are both null and timelike geodesically complete. We also
applied the counterterm method to the case of our solutions in dimensionally
continued gravity and calculated the finite mass of the spacetime. We found
that the counterterm (\ref{Ict}) has only one term, since the boundaries of
our spacetimes are curvature-free. Other related problems such as the
application of the counterterm method to the case of solutions of Lovelock
gravity with nonzero curvature boundary remain to be carried out. Also it
would be interesting if one can generalize the static uncharged solution
introduced in this paper to the case of static or rotating charged solutions
of Lovelock-Maxwell gravity.

\acknowledgments {This work has been supported by Research
Institute for Astrophysics and Astronomy of Maragha, Iran.} 

\end{document}